\documentclass[sn-basic]{sn-jnl}

\usepackage{graphicx}%
\usepackage{multirow}%
\usepackage{amsmath,amssymb,amsfonts}%
\usepackage{amsthm}%
\usepackage{mathrsfs}%
\usepackage[title]{appendix}%
\usepackage{xcolor}%
\usepackage{manyfoot}%
\usepackage{algorithm}%
\usepackage{algorithmicx}%
\usepackage{algpseudocode}%
\usepackage{subcaption}
\usepackage{url}
\usepackage{hyperref}

\begin{document}


\title[Interpretable Model for Investment Decisions]{Enhancing Profitability and Investor Confidence through Interpretable AI Models for Investment Decisions}



\author[1,2]{\fnm{Sahar} \sur{Arshad}}\email{sarshad.phdcs20seecs@seecs.edu.pk}

\author*[1]{\fnm{Seemab} \sur{Latif}}\email{seemab.latif@seecs.edu.pk}

\author[1]{\fnm{Ahmad} \sur{Salman}}\email{ahmad.salman@seecs.edu.pk}

\author[3]{\fnm{Rabia} \sur{Latif}}\email{rlatif@psu.edu.sa}

 \affil[1]{\orgdiv{School of Electrical Engineering and Computer Science}, \orgname{National University of Sciences and Technology (NUST)}, \orgaddress{\city{Islamabad}, \postcode{44000}, \country{Pakistan}}}

 \affil[2]{\orgdiv{Department of Computer Science}, \orgname{Bahria University}, \orgaddress{\city{Islamabad},  \country{Pakistan}}}

\affil[3]{\orgdiv{College of Computer and Information Sciences (CCIS)}, \orgname{Prince Sultan University}, \orgaddress{\city{Riyadh},  \country{Saudi Arabia}}}



\abstract{Financial forecasting plays an important role in making informed decisions for financial stakeholders, specifically in the stock exchange market. In a traditional setting, investors commonly rely on the equity research department for valuable reports on market insights and investment recommendations. The equity research department, however, faces challenges in effectuating decision-making due to the demanding cognitive effort required for analyzing the inherently volatile nature of market dynamics. Furthermore, financial forecasting systems employed by analysts pose potential risks in terms of interpretability and gaining the trust of all stakeholders. This paper presents an interpretable decision-making model leveraging the SHAP-based explainability technique to forecast investment recommendations. The proposed solution not only provides valuable insights into the factors that influence forecasted recommendations but also caters the investors of varying types, including those interested in daily and short-term investment opportunities. To ascertain the efficacy of the proposed model, a case study is devised that demonstrates a notable enhancement in investor's portfolio value, employing our trading strategies. The results highlight the significance of incorporating interpretability in forecasting models to boost stakeholders’ confidence and foster transparency in the stock exchange domain.}




\maketitle

\section{Introduction}\label{sec1}

Decision-making in financial markets is a complex cognitive process involving the analysis of various options to choose the most optimal course of action. In the rapidly evolving landscape of financial markets, this process has become increasingly significant, especially within the volatile environment of the stock market. Investment decisions in the stock market require navigating an intricate web of fundamental analysis, market trends, economic indicators, and market sentiments \citep{frydman2016psychology}. Additionally, recognizing and promptly capitalizing on opportunities is essential for investors to leverage short-term market fluctuations in their strategies. Given the dynamic nature of the stock market, time-sensitive, informed investment decisions towards buying and selling securities facilitate distinct market participants.

\textbf{Investment Challenges and Strategic Practices in Equity Research}

In a conventional stock market environment, the equity research division fosters a collaborative and informative relationship with market participants. Through regular research reports, trend forecasts, and personalized recommendations, the research team provides valuable insights into potential market opportunities. This involves adopting various analytical methods and essential forecasting to ensure that informed decisions align with market participants’ investment goals \citep{thakkar2021fusion}. However, the equity research division faces significant obstacles in publishing timely research reports due to the extensive cognitive effort required to assess and synthesize the sheer volume of diverse market data. This process can lead to biased decisions from analysts with different backgrounds and experiences. Additionally, frequent price fluctuations necessitate timely investment recommendations for different investor groups \citep{dolatsara2022interpretable}. Each group has unique needs based on their investment strategies, requiring a nuanced approach to provide relevant and actionable insights. For active traders, the analysis must highlight real-time market trends and specific entry/exit trading points. Weekly traders may seek more comprehensive analysis, including longer-term trends and preferences for certain technical indicators. The integration of automation into traditional research processes empowers analysts to customize forecasted recommendations based on individual investor preferences, thereby enhancing decision-making across different segments promptly.

\textbf{Emergence of Explainable AI (XAI) in Financial Forecasting}

In recent years, the domain of financial forecasting has garnered considerable attention from the research community \citep{nti2020systematic}. This interest is driven by advancements in Artificial Intelligence (AI) and the abundant availability of financial data in digital form. AI plays a significant role in deploying financial services in various industries, including trading financial instruments \citep{huang2019automated}, creditworthiness profiling \citep{bussmann2020explainable}, credit or loan decision-making \citep{misheva2021explainable}, fraud detection \citep{soviany2018benefits}, stock price prediction \citep{li2020incorporating}, insurance claim assessment \citep{riikkinen2018using}, and debt collection management \citep{phillips2019artificial}. Utilizing an automated forecasting system with intelligent investment recommendations is indispensable for making profitable decisions in the dynamic and complex financial market \citep{moșteanu2019international}. However, the adoption of such AI-based solutions faces significant challenges, especially in fields like health, defence, and finance, where decision-making can impact well-being. Regulatory authorities, such as the European Union, have expressed concerns about the inability to interpret and explain the decision-making processes underlying complex AI-based black-box models \citep{arrieta2020explainable}. Given the highly regulated environments in which financial intermediaries operate, it is crucial to deploy explainable modules for large-scale solutions. In time-critical scenarios, the equity research department's responsibility towards diverse investors underscores the necessity for transparent and comprehensible methodologies, instilling confidence in those who rely on their reporting for investment decisions.

Recently, Explainable AI (XAI) has emerged, adding explainability and understandability to the outcomes of deployed automated models without compromising performance and accuracy. While financial forecasting has traditionally been treated as an AI problem \citep{carta2021explainable}, there are few studies \citep{gite2021explainable,xing2019sentiment, yang2018explainable} that have integrated transparent and trustworthy AI-based financial solutions. Addressing this research gap, this study focuses on exploring and applying a transparent and automated forecasting system to ensure adherence to regulatory standards within the financial sector.

\textbf{Study Contributions}

This study makes several key contributions to addressing and incorporating interpretable decision-making in the context of investments in the Pakistan Stock Exchange (PSX):
\begin{enumerate}
    \item Stakeholder Identification and Needs Assessment: The proposed forecasting model considers and implements variations among different types of investors, addressing their unique needs and rights.
    \item Context-Specific Analysis: Unlike international stock exchange benchmarks such as S\&P 500 and DJIA, PSX operates within a distinct context influenced by factors like foreign investments, and political and economic stability, rather than global economic indicators like CPI and unemployment rates.
    \item Quantifying Decision-Making Factors: The study involves quantifying and modeling the factors influencing human economic decision-making, using technical indicators to generate sound reports that help users understand the rationale behind the model's forecasts.
\end{enumerate}

The subsequent sections of the paper are structured in the following manner: Section 2 presents a review of notable research studies that have investigated the intersection of financial forecasting and explainability within the context of financial markets. This is followed by Section 3, which presents our proposed methodology. It integrates the explainability components in the solution addressing the requirements of different types of investors that can be addressed via the equity research department. Section 4 provides the results and discussion as well as a case study that demonstrates the implementation of the proposed methodology. Finally, Section 5 concludes and highlights the paper's findings along with the future direction of this research effort.

\section{Literature review}\label{sec2}

In recent years, several research studies on financial forecasting have been conducted, primarily aimed at easing the decision-making process for investors. This literature review places special emphasis on the efforts of researchers to provide explanations within the context of financial markets, specifically concerning the decision-making process resulting from proposed solutions. As opined by researchers \citep{ballinari2022does,ben2022dynamic}, the complexity and unpredictability of financial markets make it challenging to accurately forecast future trends. Consequently, there has been an increasing interest in incorporating explainability into forecasting models to enhance transparency and interpretability for investors in the decision-making process.

The stock market is a complex system that is influenced by numerous factors, including economic indicators, political events, and investor sentiment. \citep{li2020incorporating} developed a prediction system for stock price return rates that integrates both technical and sentiment analysis. Their approach enabled the prediction model to learn sequential information intelligently within a market snapshot time series. The objective was to classify the stock trends based on the market information that was integrated with textual content from the news domain along with the current circumstances in the stock market. This has been achieved by constructing a two-layer LSTM neural network to learn the sequential information followed by a fully connected network to make stock predictions. To assess the efficacy of the proposed method, the researchers constructed a system using five years of daily stock price data from the Hong Kong Exchange (HKEx) for four prominent sectors (Commerce, Finance, Properties, and Utilities) and incorporated the corresponding FINET news articles from the same period, along with four sentiment dictionaries. Through a series of experiments, it was found that among the four sentiment dictionaries, the finance domain-specific sentiment dictionary (Loughran–McDonald Financial Dictionary) models the news sentiments better, which brings more prediction accuracy and F1 Score than other baseline approaches. Although the sentiment space constructed by utilizing the dictionaries shows improvement in the stock price prediction at individual stock, sector, and index levels, however, the proposed approach did not consistently outperform baseline approaches in all sectors of the stock market and this needs to be further investigated. Additionally, it is found that despite the availability of longer-term data, the trading history is limited to a shorter period. Yet, strong indicators like EMA200 or EMA100 are widely adopted by regular financial analysts in their technical perception and can also be effectively utilized in conjunction with appropriate explanations about the model's predictions.

In another study, \citep{carta2021explainable} presented an NLP-based approach for financial forecasting keeping in view the exogenous factors that affect the stock prices in a financial market. The goal was to achieve a methodology, both competitive in forecasting performance and explainable in the obtained output for the individual companies listed in the S{\&}P 500 index. In the procedure, industry-specific lexicons were generated from the Dow Jones ‘Data, News, and Analytics (DNA) dataset. These lexicons were then refined through feature engineering, which involved selecting impactful words from the articles to capture certain statistical indicators associated with a company in a given time. The resulting list of features was used to train a decision tree classifier to predict whether a company's stock price listed in the S{\&}P 500 Index will rise or fall on the following day. Finally, the model explanation was achieved by presenting the rules to the users that are determined by the predictions drawn from the classifier. The explanation was complemented by a list of sentences, containing relevant lexicon words, selected from the groups of news articles considered in the feature extraction phase. The proposed model not only achieved a better F1 score in prediction but also may function as a superior tool for visualizing and explaining the results in the form of boolean logic conditions. However, their proposed solution falls short in incorporating technical indicators into the system to generate reports for end-users. To provide sound explanations, analysts must comprehend other technical insights beyond news, since mere words from news do not substantially alter the fluctuations in the stock market. Moreover, there is a need to cover the semantic relationship between the aggregated news over time for more accurate and comprehensive results. 

\citep{theodorou2021ai} developed a platform called ASPENDYS to analyze daily stock data based on technical analysis as well as to consider news text reliability and social media sentiment concerning stocks. They provided a high-level interactive interface offering end users the functionalities of monitoring, modifying, and expanding their portfolio. The aim was to maximize the expected return of a portfolio by capturing the sentiment regarding a specific asset listed on the Toronto Stock Exchange (TSX60). With the proposed architecture, an efficient portfolio management system was developed where different investment strategies were adopted depending on the investor type. The pipeline consisted of seven components that collected data from different sources, including social media (Twitter and Stocktwits), news articles, and financial data. The news data was checked for reliability using a Random Forest classifier, trained on the CREDBANK dataset and sentiment analysis was performed using BERT, fine-tuned on the Stanford Sentiment Treebank-5 (SST-5) dataset. Investment signals were generated using four sub-modules based on different input data, including sentiment, financial, and machine learning signals. Finally, the user portfolio is optimized using Modern Portfolio Theory, Optimized Portfolio, Black-Litterman model, and Total Variation. As an experiment, two portfolios were constructed to represent different types of investors. These portfolios were designed to illustrate the contrast between the investment history of signals over time that were accepted versus those that were ignored, depending on the type of user. Although the need for explainability is highlighted, it is not addressed explicitly in the research.

\citep{yang2020generating} investigated the production of post hoc plausible counterfactual explanations for an efficient financial solution using a robust transformer model. The objective of the study was to create an investment approach with significant returns by speculating on financial applications in the Fintech field. The authors of the study collected their dataset from Zephyr, a comprehensive database of real-world deal data, and were able to qualify over 4000 instances after preprocessing. Afterwards, the authors utilized three methods - REP-SCD, RM-SCD, and INS-SCD to perform their prediction task using a classical transformer architecture on a given news article. As a next step, the explanation generated for the present prediction was flipped with appropriate words utilizing the three techniques that changed the prediction class of the original input. The final step involved generating counterfactual explanations using substitutions of grammatically plausible words. The results demonstrated superior accuracy and explanatory performance as compared to state-of-the-art techniques. The qualitative assessment of the more grammatically plausible explanations was evaluated by financial experts which received a score of 0.85. Although the authors focused on predicting mergers and acquisitions in financial companies with their proposed framework, however, they did not test it for cancelled or pending deals. Additionally, relying solely on the positivity of actual news to determine the likelihood of a deal being completed may not adequately capture the complexities of real market dynamics \citep{theodorou2021ai}.

In another notable effort from researchers, aiming to provide explainable stock movements, \citep{gite2021explainable} applied a fusion of news headlines sentiments and financial data utilizing the state-of-the-art deep learning model Long short-term memory combined with a convolutional neural network (LSTM-CNN), for the prediction task. Followed by integrating a model-agnostic explainable module using Local Interpretable Model-Agnostic Explanations (LIME) to interpret the model prediction, where input features from Sensex news headlines were given weights in explaining the predicted outcome. Although the accuracy reported is 96.2\% and contextual information is also extracted from the textual data by assigning weights to important news content. However, the explanation module does not represent the context adequately and only provides word vectors as an outcome, indicating the need for more convincing explanations to users.

\citep{xing2019sentiment} in continuation to their prior work \citep{luo2018neural}, incorporated explicit knowledge in the proposed sentiment-aware volatility forecasting (SAVING) model to forecast stock returns volatility. A knowledge-based approach that employed SenticNet was utilized in acquiring sentiment polarity from the social streaming on StockTwits for 10 US hotspot stocks. The proposed framework utilized the modern way of pricing derivatives such as the Black-Scholes model for stock volatility. A variety of models equipped with sentiment information from benchmark groups including linear and deep models were evaluated. Experiments showed the SAVING model outperformed many state-of-the-art volatility forecasting models including GARCH, VRNN, and NSVM. Even though the naïve generalized model has the expressive power of deep VRNNs and can capture bidirectional interaction between market sentiment and stock prices, the authors plan to enhance their approach by utilizing sentiment knowledge bases that are specific to the finance domain. Furthermore, the explainability module used in this study is limited to analyzing the polarity of individual sentences and requires a more sophisticated approach to be integrated to provide comprehensive explanations.

Stock market rumors refer to special information that begins to circulate in the stock market without confirmation. A new perspective of stock price linkage has been focused on by \citep{dong2019price}, highlighting the interplay between various types of investors in the stock market and how it can influence their risk preferences and behavior when it comes to rumors. To capture the complexity of the contagion process, the SCIR contagion mathematical model was constructed which incorporated the spread of risk from one investor to another, both within and across multiple networks of investors. By simulating the effects of rumors and certain probability mechanisms that can trigger risk contagion in the stock market on heterogeneous investors, the study aimed to assist regulators in devising effective strategies for mitigating financial risks. However, it is essential to conduct experiments with real market data to gain a better understanding of the emotional and psychological factors that influence investor behavior in actual trading scenarios.

\citep{yang2018explainable} suggested using a two-level attention mechanism based on a gated recurrent unit (GRU) network to forecast changes in stock prices. This approach involves analyzing the most important events from the last seven days' news from Reuters and Bloomberg and providing explanations for the predictions. The study also introduced a financial knowledge graph of entities listed in the S{\&}P to deal with information from news items that cannot be processed in the initial stage. Although, GRU-2AT outperforms two baselines without attention mechanisms in terms of accuracy and Matthews Correlation Coefficient (MCC) score for three companies from different regions. However, the study has limitations such as the dependence on news only for stock price prediction and considering only three S{\&}P 500 listed companies for the knowledge graph. Additionally, there is no visualization for the explainability module of the GRU model about the predictions made.

\citep{Carta2021ExplainableAF} highlighted the significance of employing explainable AI techniques in financial forecasting. They specifically addressed the challenge of low correlation between the extracted financial features and the target variable. The authors utilized machine learning algorithms to automatically select relevant features. The XAI methods were then utilized to explain the importance and relevance of the selected features in the context of financial forecasting. This increases the predictive performance of the model by identifying and removing uninformative features by identifying optimal thresholds for each stock in the heterogeneous stock set. The proposed permutation-based strategy is compared to the state-of-the-art approaches i.e., Random Forest feature importance and Local Interpretable Model agnostic Explanations models (LIME). Although the authors have emphasized the perspectives of the explainees or recipients of the explanations, there is a lack of a mechanism in place to identify and manage various stakeholders in StatArb financial solution\citep{carta2022explainable}.

There are several key limitations in the existing research that need to be addressed, including but not limited to the insufficient adoption of popular technical indicators that are usually exercised during the technical or fundamental analysis for financial decision-making by the market stakeholders. Moreover, there is a lack of explicitly addressing varying stakeholders’ requirements while providing them with explanations. In a few studies, there is a gap in conducting experiments with real market data which is better for comprehending the investor's behaviour and thus provisioning them with interpretations in a context. By addressing these gaps, the research community can establish a potential bridge between academic finance and the financial industry, enabling the development of more practical financial forecasting models, integrated with sophisticated explainability approaches. To the best of our knowledge, this is the first study of its kind that establishes the effort of providing visual explanations tailored to varying stock exchange investors. Moreover, the study integrates widely used technical indicators to analyze the data of Pakistan's stock exchange, adding a new dimension to the research.
\section{Methodology}\label{sec3}

In this research work, we acquired business-day financial data from the official website of the Pakistan Stock Exchange \citep{CiteDrive2022} as well as \citep{CiteDrive2023}. Our data spanned over 10 years, specifically from 2012 to 2022, comprising approximately 2400 index entries, ensuring that we capture enough sample size to achieve our research objectives. The collected data encompasses various attributes of the KSE-100 Index, including the opening (O), highest (H), lowest (L), and closing (C) prices, as well as the traded volume (V) and the change in closing price $(\Delta C)$. These predictors provide the identification of trends and potential price reversals in the stock market which are crucial in making informed investment decisions. We can represent our raw feature space, $F_d$ as follows:

Let, $F_d = \{O_d, H_d, L_d, C_{d}, V_d, \Delta C_d\}$ for a period represented as $d = 1, 2, 3, \ldots, D$  where \textquotesingle d\textquotesingle represents a specific day within the time period \textquotesingle D\textquotesingle  which in turn denotes the total number of the time period, corresponds to 10 years. Considering that there are approximately 250 trading days in a year, \textquotesingle D\textquotesingle can be estimated as D $\approx 10 \text{ years} \times 250 days/year$. Furthermore, the change in closing price for each day, $\Delta C_{d}$ is calculated as the difference between the closing price on the day \textquotesingle d\textquotesingle and the previous day i.e., $\{C_d - C_{d-1}\}$.

After collecting data, preprocessing and required numerical conversions were performed. To ensure the completeness of the data, missing values were handled using a forward fill method, whereby the last observed value was replaced by the missing values since this is a reasonable assumption based on the stability of stock prices from day to day. 

Based on the given stock market predictors in $F_d$, this study employs an experimental research approach by using various statistical methods as well as machine learning approaches. The statistical methods include technical analysis techniques on the raw stock data to formulate support and resistance trading strategies using indicators such as Exponential Moving Averages (EMAs), Moving Average Convergence Divergence (MACD), and Relative Strength Index (RSI). We incorporated these indicators with the curated data since these are commonly used by stock market stakeholders in their decision-making processes. Each technical indicator is presented in the following subsections for the current financial day \textquotesingle d\textquotesingle:
\begin{enumerate}[2.]
\item \textbf{EMA}

Exponential moving averages for a period of 55, 100, and 200 closing indices are applied to identify trends from shorter to longer spans in stock data. Eq. 1 represents its mathematical formulation:
\begin{equation}
    \text{EMA}_d = C_d \ast \alpha + (1-\alpha) \ast \text{EMA}_{d-1}
\end{equation}
where:
\begin{itemize}
\item $C_d$ is the KSE-100 index close price
\item ${EMA}_{d-1}$ is the EMA value for the previous day
\item $\alpha$ is a smoothing factor that determines the weight given to the most recent price versus the previous EMA value. It is calculated as $\alpha = \frac{2}{x+1}$, where $x \in \{55, 100, 200\}$ financial days.
\end{itemize}

\item \textbf{MACD}

The MACD indicator is used to identify potential changes in the trend. Eqs. 2 \& 3 represent its mathematical formulation.

\begin{equation}
\text{MACD\_Line}_d = \sum_{i=d-k}^{d} \left(\frac{\text{EMA}_i}{k}\right) - \sum_{j=d-l}^{l} \left(\frac{\text{EMA}_j}{l}\right) \tag{2}
\end{equation}
\begin{equation}
\text{Signal\_Line}_d = \sum_{m=d-9}^{d} \left(\frac{\text{EMA}(\text{MACD\_Line}_m)}{9}\right) \tag{3}
\end{equation}
where:
\begin{itemize}
	\item k and l are constants with values usually set as 12 and 26 financial days respectively.
\end{itemize}
\item \textbf{RSI}

The RSI is used to measure the momentum of the stock price and identify potential overbought or oversold conditions. Eqs. 4-7 represent its mathematical formulation:
\begin{equation}
\text{Avg.Gain}_d = \sum_{r=d-14}^{d} \left(\frac{\text{Gain}_r}{14}\right) \tag{4}
\end{equation}
\begin{equation}
\text{Avg.Loss}_d = \sum_{s=d-14}^{d} \left(\frac{\text{Loss}_s}{14}\right) \tag{5}
\end{equation}
\begin{equation}
\text{RS}_d = \frac{\text{Avg.Gain}_d}{\text{Avg.Loss}_d} \tag{6}
\end{equation}
\begin{equation}
\text{RSI}_d = 100 - \left(\frac{100}{1 + \text{RS}_d}\right) \tag{7}
\end{equation}

where:
\begin{itemize}
	\item $\text{Gain}_r$
 and $\text{Loss}_s$ are constants representing positive and negative changes in close price over financial days respectively.
\end{itemize}
\end{enumerate}

As a next step, a strategy set, $S = \{S_{\text{EMA}_x}, S_{\text{MACD}}, S_{\text{RSI}}\}$, where $x \in \{55, 100, 200\}$ for each technical indicator as mentioned above was formulated. The purpose of this strategy set is to detect possible buy and sell signals for trading based on support and resistance levels. To achieve this, we determined the optimal thresholds for each indicator by using a combination of backtesting and relative adjustment to generate trading signals of varying levels. These signals range from weak to strong buy/sell signals as well as a signal to hold the stock. The strength of the signals was determined by the threshold values, with higher threshold values indicating a stronger signal. The generated signals are not only helpful in making informed decisions for traders but also in quantifying the factors influencing their investment decision-making. For visualization and analysis purposes, each strategy result was plotted on the chart to assess the individual impact on the price chart of stock data. This enabled us to determine the contribution of each strategy in formulating our target feature of the dataset. 

To formulate the target feature, $Z_t$ we used a grid search technique to find optimal weights for each technical indicator-based strategy in S. A dictionary of weights W, is defined which contains the range of weights, denoted as $w_i for \textit{i} \in \{1, 2, 3, 4, 5\}$ for each strategy. We then selected the set of weights with the highest accuracy as the overall optimal weights. A weighted voted function is then utilized with these optimal weights for each strategy to formulate a final trading signal for the investor for the following day. 

The weighted vote function is expressed in Eq. 8:

\begin{equation}
    wv(S,W) = \text{sign}\left(\frac{\sum (w_i \ast s_i)}{\sum w_i}\right) \tag{8}
\end{equation}

where:
\begin{itemize}
    \item wi is the $i^{th}$ weight in W
    \item $s_i$ is the $i^{th}$ signal in S
    \item $\sum$ denotes the sum of elements in a vector
    \item $sign$ returns the polarity of the result; either buy or sell
    \item wv (y, w) gives output to a single signal (3 for strong buy, -3 for strong sell, 2 for moderate buy, -2 for moderate sell, 1 for weak buy, -1 for weak sell, and 0 for hold)
\end{itemize}

To validate our result, two investment strategies were designed, namely scalping and hold-and-wait to address the two types of investors i.e., daily active traders and weekly investors respectively. As for scalping, the accuracy of the one-day forecast is calculated by comparing the predicted target value with the actual price change for the following day from the curated stock data. The resulting target column, $Z_{t+1}\in [-3,3]$ is then used as a basis for making active trading investment decisions. Whereas, in a hold-and-wait investment strategy, the accuracy of predicted target values is assessed by comparing them with the actual price changes occurring over the following week. The resulting target column, $Z_{t+5} \in[-3,3]$ is then used in the facilitation of short-term investment decisions by the investors. Algorithm 1. presents the pseudocode for the later investment strategy.
\begin{algorithm}[H]
\caption{Trading investment strategy for short-term investor class}\label{algo1}
{\bfseries Input}:
\begin{itemize}
\item $D \Leftarrow \{ \text{ema55}_d, \text{ema100}_d, \text{ema200}_d, \text {macd}_d, \text{rsi}_d \}$ and $d=1,2,3……D$ where \textquotesingle d\textquotesingle represents a specific day over a period \textquotesingle D\textquotesingle
\item Strategy set S, representing trading signals
\item A dictionary of weights W
\end{itemize}

{\bfseries Output}:
\begin{itemize}
\item $Z_t+5$ as target column in D
\end{itemize}
{\bfseries Steps}:
\begin{enumerate}
    \item Initialize a list of possible weight combinations, \(W_c = S \times W = \{(s, w) \, | \, s \in S, w \in W\}\) is Cartesian product of two sets S and W.
    \item 	Iterate over each set of weights in $W_c$, and do the following:
    \begin{itemize}
        \item Calculate a weighted vote function, $wv(D,w_{cj} )$ as in Eq. 8 for each row in D with $w_{cj},$ where $w_{cj} \in W_c$  to obtain $Z'_{t+5}$ as intermediary target column formulation.
        \item Call the $Agg\_signal\_5dayForecast$ function to calculate the accuracy for the weekly forecast using 20\% of $Z'_{t+5}$ and comparing with the next week’s change in closing price $\Delta C_{d+5}$  for $\Delta C_{d+5} \in F_d$ where \textquotesingle d\textquotesingle represents a specific day within the time period.
    \end{itemize}
    \item 	Repeat step 2 for each weight combination $w_{cj}$ to get the best accuracy corresponding to the best weights $w_{best}$ in formulating optimized $Z_{t+5}$.
    
\end{enumerate}
\end{algorithm}

Finally, our proposed approach incorporates the SHAP (SHapley Additive exPlanations) module to interpret and visualize the decisions of the statistically weighted vote function for each of the investment strategies outlined above. SHAP is a model-agnostic method for explaining model predictions and is based on based on cooperative game theory \citep{lundberg2017unified}. It is a popular method for understanding the contribution of each input feature to the output of a model for a better understanding of a model’s behavior and reasoning. In our case, the SHAP module provides a framework for interpreting the contribution of each strategy based on technical indicators in generating the final buy/sell signal and can be used to quantify which indicators are most influential in the decision-making process. This allows increasing transparency and accountability of the model's decision-making process by providing a clear and interpretable explanation, which can help investors make more informed investment decisions. Within SHAP library, two classes were used namely, Kernel Explainer and Tree Explainer based on the choice of the model in hand either statistical or machine learning based respectively. Kernel Explainer utilizes a background dataset to estimate the expected model output and computes the SHAP values based on this estimation whereas Tree Explainer takes advantage of the tree structure to efficiently compute the SHAP values for tree-based models.

We have compared our results with other state-of-the-art baseline approaches i.e., Random Forest and XGBoost to evaluate its performance and accuracy in forecasting the financial data of PSX, as well as its interpretability and explainability. The selection of these machine learning algorithms is motivated by the empirical findings, presented in the studies \citep{huck2019large,krauss2017deep}, which demonstrate their effectiveness in producing valuable trading signals for portfolios characterized by frequent turnover and short holding periods over dual forecasting horizons. Given the dataset, $F_d$ along with the technical indicators, feature engineering, followed by preprocessing of the data was performed in preparing the input for Random Forest Classifier. Heatmaps were utilized to identify and then eliminate the highly correlated features in the dataset. To avoid multicollinearity and redundant information, we removed $\Delta C_d$ i.e., the change in closing price each day which was overlapping with the $C_d,$ the closing price. Furthermore, to avoid any feature domination, feature normalization was employed on the final dataset.

To strengthen our experimentation, we have also devised a case study that demonstrates how automating the short-term investment strategy using the generated investment signals can lead to increased gains in an investor's portfolio. However, it is important to note that this study focuses on daily active to short-term investor strategies and does not cover long-term investor strategies. Therefore, the findings and recommendations of this study may not apply to long-term investment horizons or strategies
\section{Results and discussion}\label{sec4}
\subsection{Experimental setup}

We conducted our experiments using Google Colab, utilizing Python version 3.10. Several packages were utilized including but not limited to Numpy, Pandas, and Scikit-learn for numerical computations, data analysis, machine learning algorithms, and evaluation metrics. The pandas\_ta library, which is a popular Python package, is used for technical analysis in financial markets. It offers a wide range of technical indicators, making it convenient for analyzing and processing financial time series data. To train the Random Forest Classifier and XGBoost models, the dataset was partitioned into training and testing sets, maintaining an 80-20 ratio. To assess model performance and generalization capabilities, a cross-validation approach was adopted, employing 5-fold splitting. Accuracy scores were computed for gauging the model's performance across each fold.

For explainable module integration, we incorporated Kernel Explainer and Tree Explainer SHAP classes in our Python implementation. In the case of Kernel Explainer, the background dataset is created using k-means clustering on input data to create a diverse representative set of samples from each cluster where the number of clusters i.e., 50 was determined using the elbow method. On the other hand, for Tree Explainer, SHAP values for each class of the target variable were computed. This provides a mechanism to understand the individual feature impacts for each class separately.

\subsection{Feature aggregation, model forecasting and trading}

Several sets of experiments were performed to ensure result validation and minimize the impact of randomness.
\begin{figure}[H]
\centering
\includegraphics[width=1\textwidth]{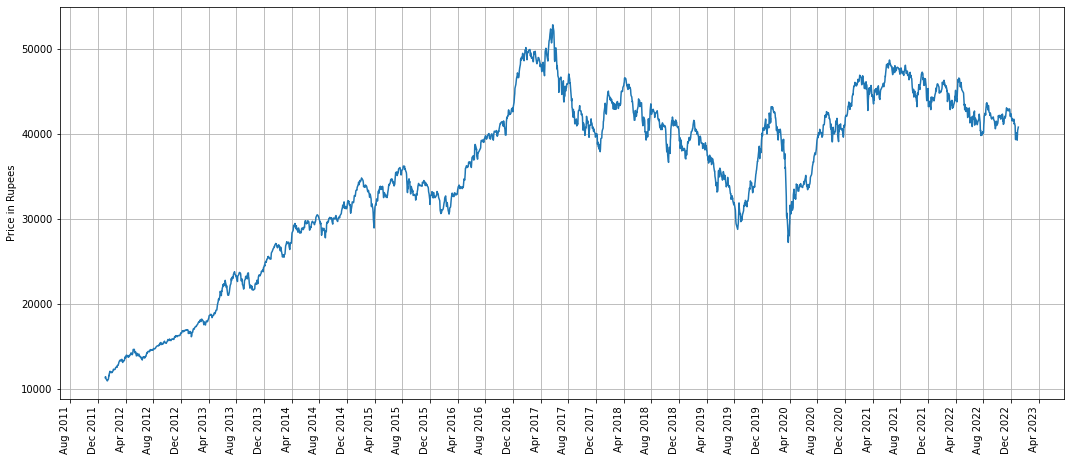}
\caption{} \label{fig1}
\end{figure} 
Fig. \ref{fig1} presents the visualization of KSE-100 closing index of the cumulative stock assets over the 10-year of period with quarterly intervals highlighting significant peaks and falls in the data.
\begin{figure}[H]
\centering
\includegraphics[width=1\textwidth]{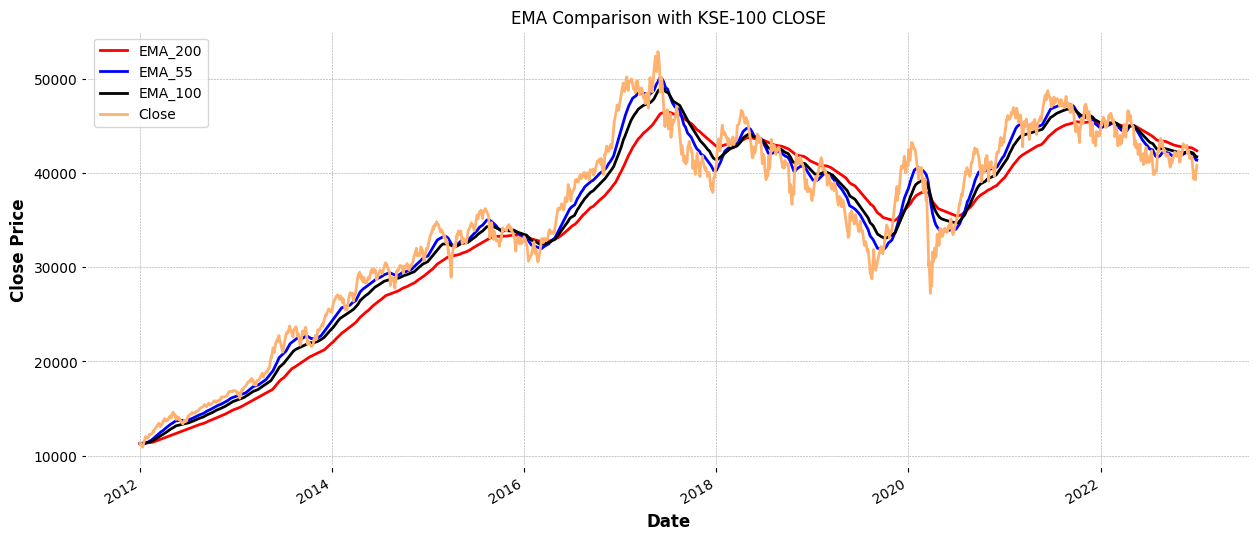}
\caption{}\label{fig2}
\end{figure}

Fig. \ref{fig2} shows various exponential moving averages along the close indices to identify short-term, medium-term, and long-term trends in the index movements respectively. This provides investors the insights into potential support and resistance levels in determining buying or selling opportunities in the stock market.
\begin{figure}[H]
\centering
\includegraphics[width=1\textwidth]{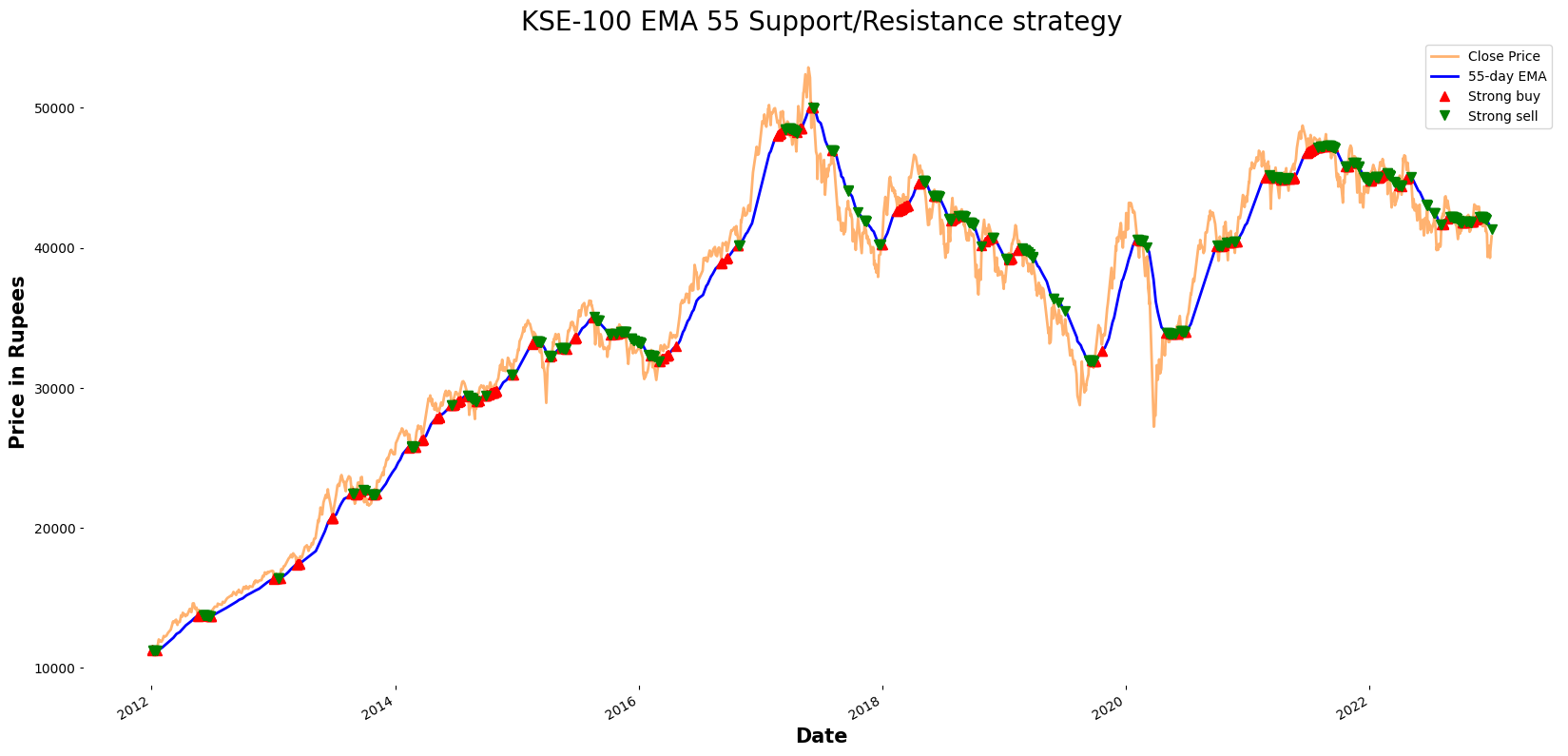}
\caption{}\label{fig3}
\end{figure}
Fig. \ref{fig3} shows the trading signals generated over time by our proposed model with the EMA-55 support/resistance strategy. It shows almost an overfitting trend fluctuates rapidly with market dynamics and generates maximum buy-sell signals, therefore, providing a resistance trend overall which is good for short investments and naive investors.
\begin{figure}[H]
\centering
\includegraphics[width=.9\textwidth]{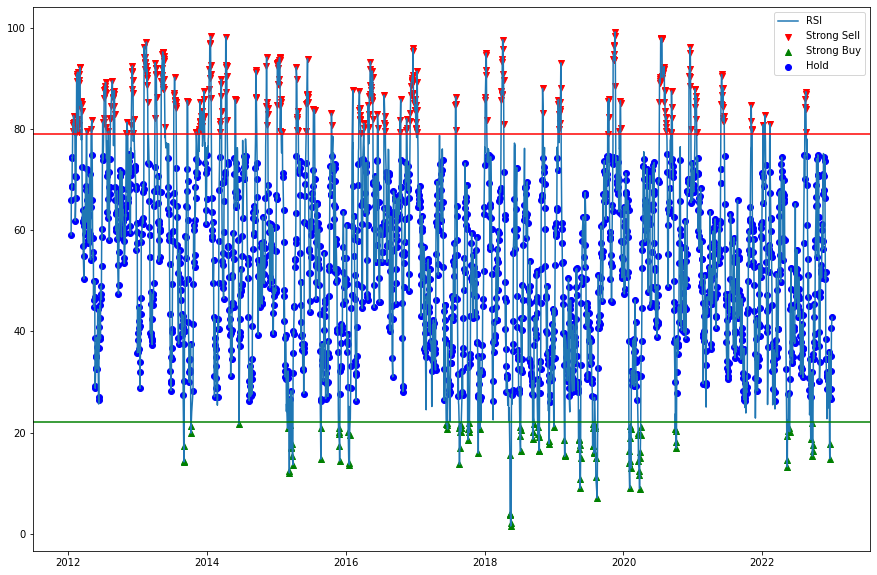}
\caption{}\label{fig4}
\end{figure}

Fig. \ref{fig4} presents the trading signals generated from the Relative Strength Index strategy, showing the over/under bought and sold conditions in the stock market. Usually, overbought levels are situated between 70 and 100 whereas oversold condition is between the levels of 30 and 0.

\subsection{Explanation for daily active trading and short-term trading strategies}

\begin{figure}[H]
\begin{subfigure}{\textwidth}
\centering
\includegraphics[width=1\textwidth]{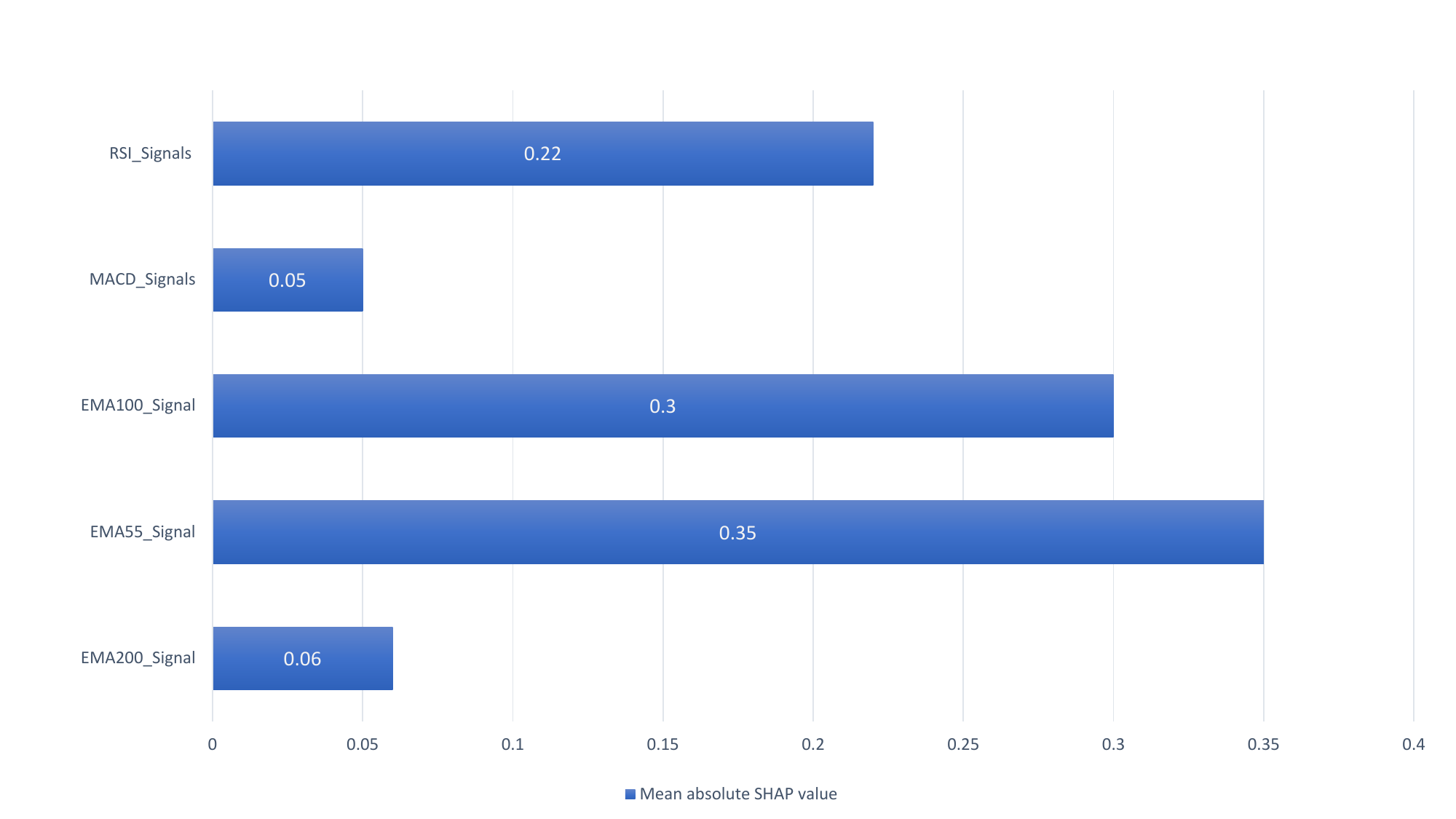}\hfill
\caption{}\label{fig5a}
\end{subfigure}

\begin{subfigure}{\textwidth}
\centering
\includegraphics[width=1\textwidth]{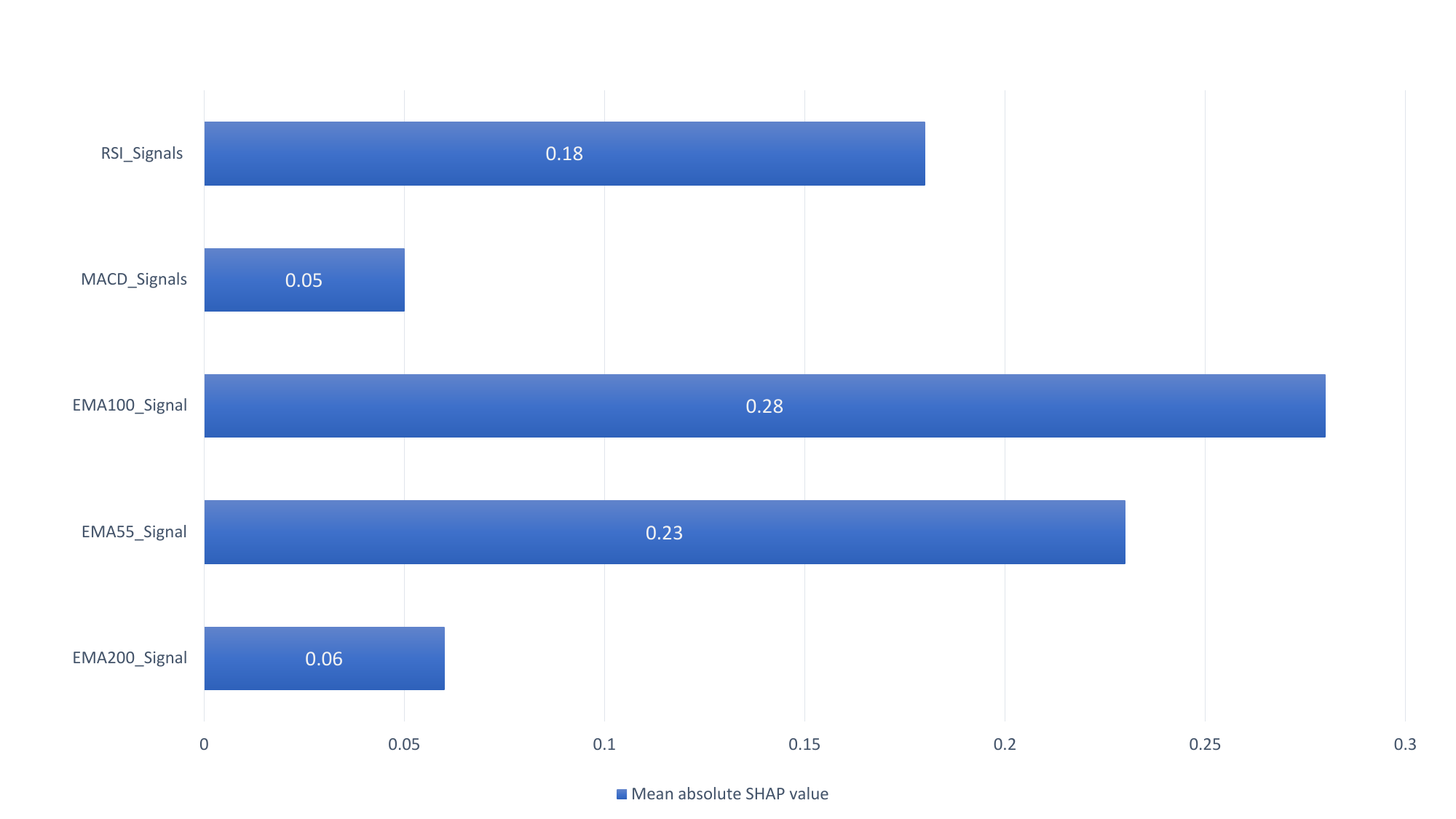}\hfill
\caption{}\label{fig5b}
\end{subfigure}
\caption{}
\end{figure}

The aggregate statistical strategy approach achieves 63.36\% accuracy by employing the optimal weights on each technical indicator i.e., EMA200\_Signal: 3, EMA55\_Signal: 2, EMA100\_Signal: 1, MACD\_Signals: 1, RSI\_Signals: 5 utilizing majority weighted voting system. Fig. \ref{fig5a} shows a visual representation of the mean SHAP values for each feature. It shows a significant impact of EMA55 followed by EMA100 and then RSI, in case of daily active trading or scalping. We can observe rare resistance and support from EMA 200 in this strategy. On the other hand, for short-term trading strategy, by employing the optimal weights i.e.,  EMA200\_Signal: 2, EMA55\_Signal: 1, EMA100\_Signal: 2, MACD\_Signals: 1,  RSI\_Signals: 2, an accuracy of 71.88\% is achieved. Fig. \ref{fig5b} shows the significant impact of EMA100 followed by EMA55 and then RSI, with aggregate weekly statistical weighted vote. This suggests that EMA100 plays an important role in this strategy. 

\newpage

\begin{figure}[H]
\begin{subfigure}{\textwidth}
\centering
\includegraphics[width=1\textwidth]{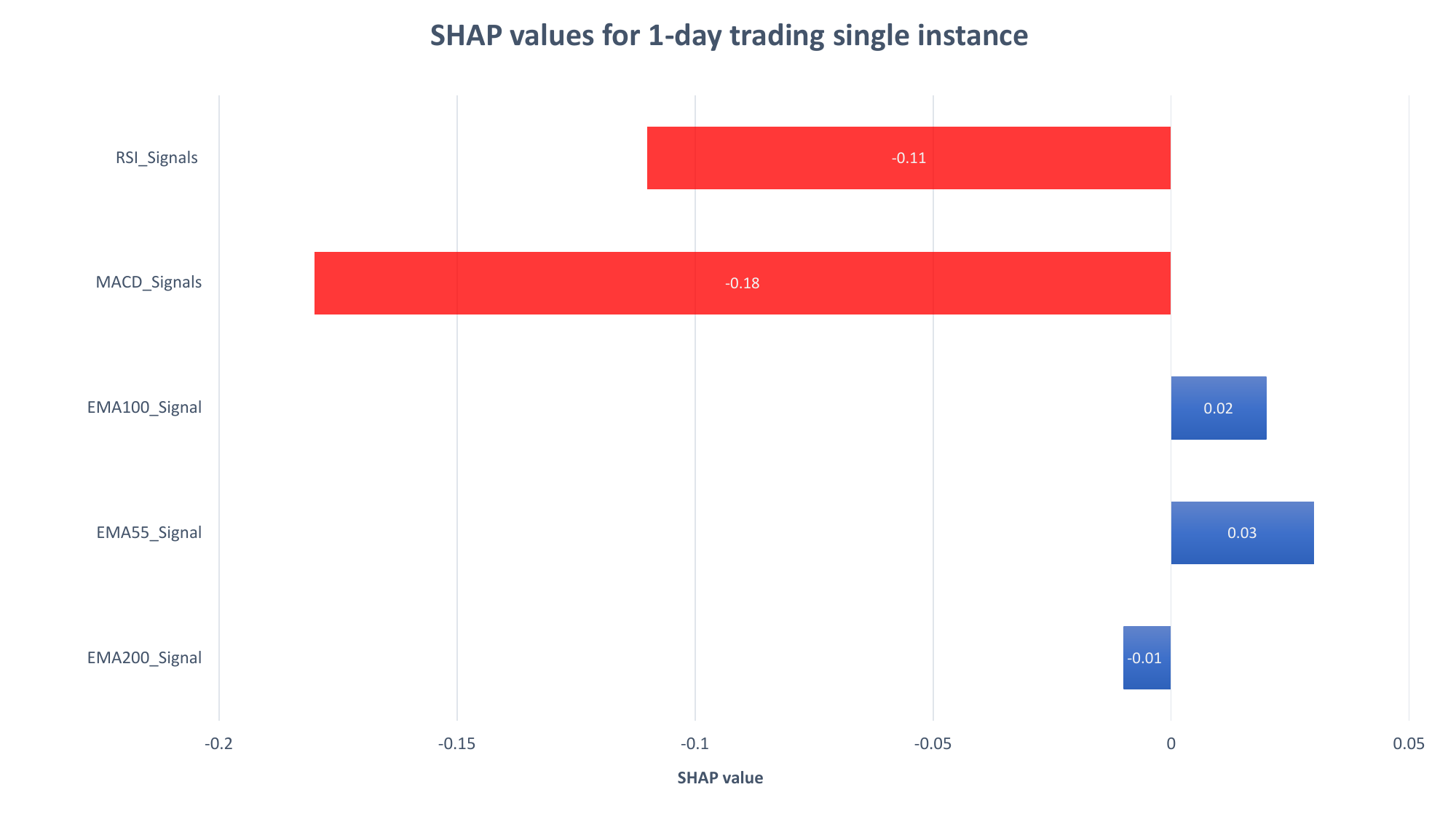}\hfill
\caption{}\label{fig6a}
\end{subfigure}
\begin{subfigure}{\textwidth}
\centering
\includegraphics[width=1\textwidth]{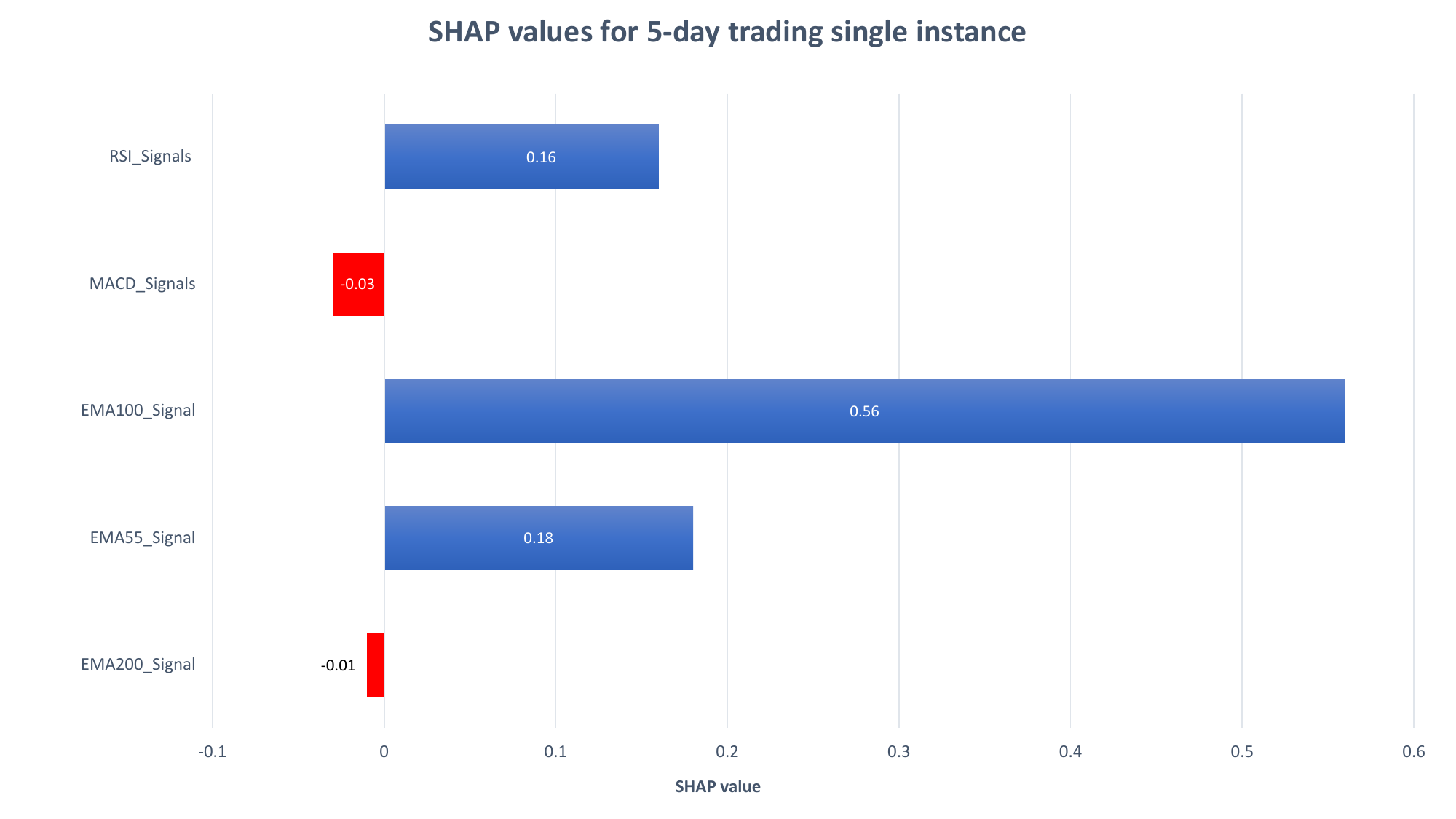}\hfill
\caption{}\label{fig6b}
\end{subfigure}
\caption{}
\end{figure}
The accuracy of a short-term trading strategy in stock investments can be attributed to the consolidation phase that occurs after market panics or significant news events. During this phase, a short-term strategy allows for market recovery or adjustment to take place, resulting in more fruitful outcomes compared to daily active trading strategies.

Fig. \ref{fig6a} and \ref{fig6b} show the block plots for daily active trading and short-term trading strategies respectively. These provide valuable insights to the traders by not only showing the influential features and polarity of a particular instance but also revealing the interaction of indicators by showcasing overlapping features contributing to the predictions. Moreover, the interpretability of plots is verified, as the SHAP values are highly correlated with the outcomes achieved through statistical strategies for 1-day and 5-day trading.

\begin{figure}[H]
\begin{subfigure}{\linewidth}
\centering
\includegraphics[width=1\linewidth]{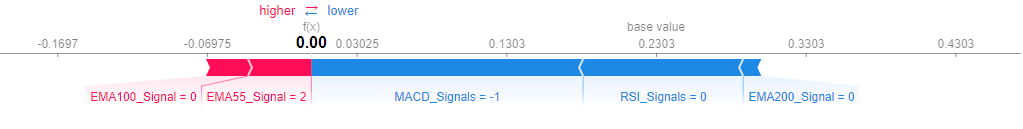}\hfill
\caption{'\textquotesingle hold a share\textquotesingle{} trading move}\label{fig7a}
\end{subfigure}
\begin{subfigure}{\linewidth}
\centering
\includegraphics[width=1\linewidth]{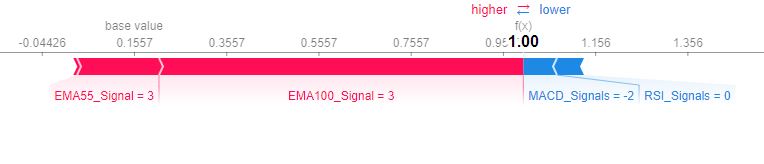}\hfill
\caption{\textquotesingle buy a share\textquotesingle trading{} move}\label{fig7b}
\end{subfigure}
\caption{}
\end{figure}

Fig. \ref{fig7a} and \ref{fig7b} show the force plots, representing instance-wise visualization for the corresponding daily active trading and short-term trading strategies. In devising a final trading move, force plots indicate the magnitude and direction of the individual feature impact on the final prediction, aiding in easy interpretation for each type of trader. We can see a decision of ‘hold a share’ for an instance in Fig. \ref{fig7a}, depicting EMA55 and EMA100 have higher contribution in the decision, when in contrast to other technical indicators. On the contrary, Fig. \ref{fig7b} depicts a trading move of ‘buy a share’ with a significant contribution of the two technical indicators.

\begin{figure}[H]
\centering
\includegraphics[width=.9\textwidth]{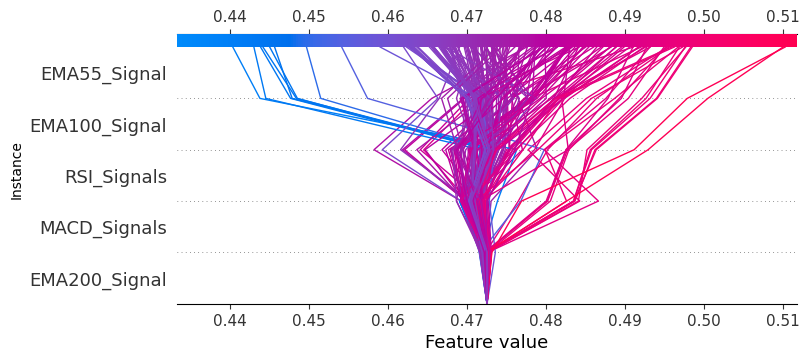}
\caption{}\label{fig8}
\end{figure}

When comparing the random forest classifier to our proposed model, it is worth noting that both models exhibited similar significant factors and contributions, as depicted in figures \ref{fig5a} and \ref{fig8} for daily active investment decisions. However, despite this similarity, the random forest classifier achieved an overall accuracy of 53\% for one day and 55\% for five days buy-sell decision forecasting. This indicates that our proposed model outperformed the random forest classifier in daily active trading and short-term investment strategies.

\begin{figure}[H]
\centering
\includegraphics[width=1\textwidth]{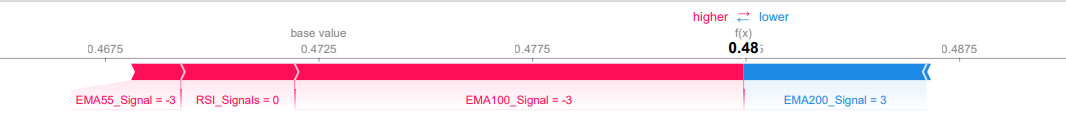}
\caption{}\label{fig9}
\end{figure}

As depicted in Fig. \ref{fig9}, a force plot is presented to illustrate a machine learning-based prediction instance. The plot highlights the influential role of EMA55, along with EMA100 and RSI, in shaping the scalping or daily active trading strategy by generating a ‘sell signal’.


\subsection{Case study of PSX stock trading}
To validate the effectiveness of the proposed trading strategy, we have devised a case study to demonstrate the application of a short-term trading strategy, using automated investments in managing a stock trader’s portfolio. The investment span is four years, starting from 2018. The initial capital investment for the portfolio is set at PKR 10,000. It is important to consider that an investor’s portfolio is practically diverse including investments in other commodities like equities and bonds \citep{jurczenko2020machine} for potential investments. When placing a ‘buy order’ as a strong signal from our strategy, the portfolio investment is utilized to buy the shares. This approach ensures that the investor capitalizes on favourable market conditions to maximize potential gains. To showcase the results of automated trading in opting for ‘sell order’, two approaches are implemented.

In the first conservative hold-and-wait approach, an investor with a stance of avoiding hasty sell decisions places a ‘sell order’ after a week’s hold of a share. In this way, the potential risks of losses are minimized and a portfolio of PKR 11,063 throughout four months of stock investment is yielded as depicted in Fig. \ref{fig10} below.
\begin{figure}[H]
\centering
\includegraphics[width=0.8\textwidth]{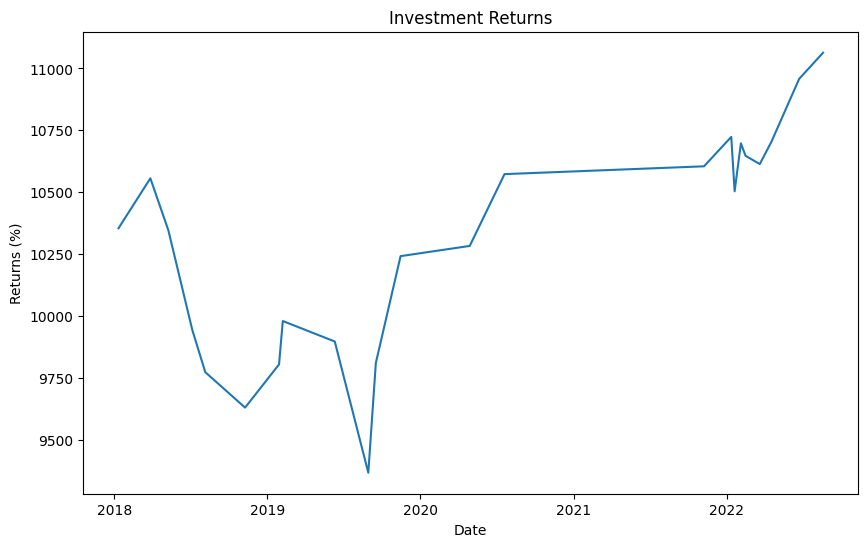}
\caption{}\label{fig10}
\end{figure}

\begin{figure}[H]
\centering
\includegraphics[width=.8\textwidth]{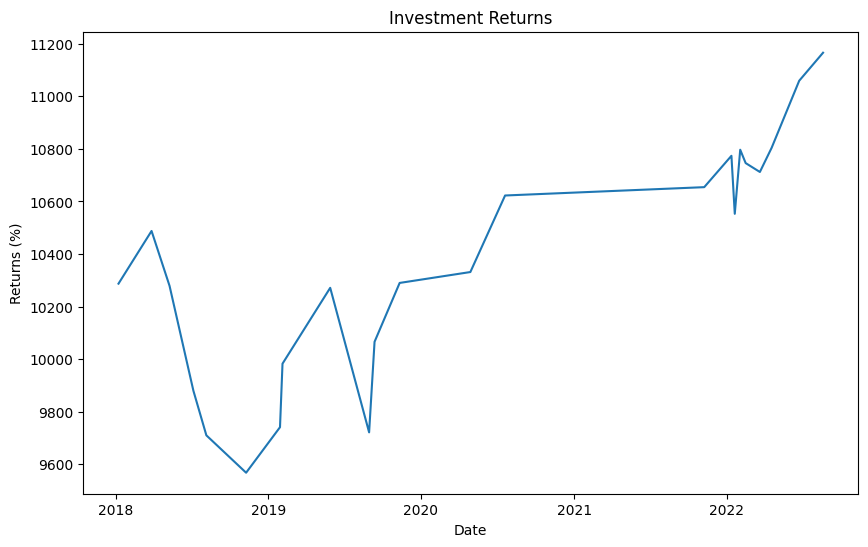}
\caption{}\label{fig11}
\end{figure}

In contrast, the aggressive investment approach places a ‘sell order’ as soon as a particular threshold of profit (i.e., 2\% or more on an original investment) is achieved. This approach depicts an optimistic investor who aims to capture gains quickly and capitalize on shorter-term market fluctuations. A better investment is yielded by utilizing this approach with a portfolio return of amount PKR 11,165. Fig. \ref{fig11} depicts the period of investments with returns over the same period as given in the first approach.
The case study demonstrates the efficacy of the proposed trading strategy in short-term stock investments. By implementing the automated trading approaches presented in this study, investors can optimize their investment decisions and potentially achieve a profit of 35\% throughout one year of stock investment which is significant as compared to other investment commodities such as equities and bonds. These findings highlight the importance of utilizing data-driven strategies and automation in stock trading to maximize returns while effectively managing risks.

\section{Conclusion and Future Work}\label{sec5}

 In this research work, we investigated an interpretable investment decision-making model leveraging the SHAP-based explainability technique. Although the proposed statistical model provides a solid foundation for inference and yields better results in terms of forecasting trading signals and enhanced interpretation, there are certain limitations inherent to such classical approaches in contrast to the state-of-the-art Machine Learning models. In real trading scenarios, market stakeholders are not merely dependent on endogenous market information. Their decision-making is also influenced by various factors, including sentiments built from news and social media, fear and greed ratios, and external economic factors such as foreign investments and exchange rates like the Dollar-to-Pakistani-Rupee rate. On the other hand, Machine Learning models, although less interpretable in contrast, can identify complex patterns and can handle data heterogeneity more flexibly. Leveraging a post hoc interpretability technique like SHAP, therefore, provides a more robust solution for ML-based black box models. In future work, we aim to employ diverse alternative data to bridge the gap between academic financial models and practical solutions. For this, a potential area of research is Natural Language Financial Forecasting (NLFF) which aims to process and interpret unstructured financial text data to generate meaningful forecasts and assist in investment decision-making \citep{xing2018natural}. It combines the power of language understanding with financial analytics to provide valuable insights and predictions in the financial domain. Another possible research direction for a deeper understanding of the market dynamics within specific sectors is to identify key drivers of stock price movements at a granular level. This could involve investigating the influence of various factors such as exogenous factors, sector-specific news, company-specific events, market sentiment, and price action patterns of individual stocks. Finally, we aim to enhance the existing explainability capability of our model by incorporating decision rules into the explanation process. In this way, a more structured and interpretable format to understand the reasoning behind the predictions is possible.

\section*{Abbreviations}
 SHAP, SHapley Additive exPlanations; AI, Artificial Intelligence; XAI, Explainable AI; PSX, Pakistan Stock Exchange; KSE, Karachi Stock Exchange; CPI, consumer price index; EMA, Exponential moving averages; MACD, Moving Average Convergence Divergence; RSI, Relative Strength Index; NLFF, Natural Language Financial Forecasting.

  \section*{Declarations}
 \section*{Availability of data and material}
The data will be available upon request. The GitHub repository \href{https://github.com/sahar-arshad/PSX-Interpretability.git}{https://github.com/sahar-arshad/PSX-Interpretability.git} contains the implementation of the algorithms and models discussed in this paper, allowing for reproducibility and further exploration of the methods used in this study.

\section*{Competing interests}
The authors declare that they have no conflict of interest.

\section*{Authors' contributions}
Sahar Arshad: Conceptualization, methodology design and implementation, along writing of the original draft.\\Seemab Latif:  Supervision, formal methodology analysis and critical review of the manuscript.\\Ahmad Salman \& Saadia Irfan: Contribution in revisions and final approval of the version to be published

\section*{Funding}
This research is funded by the Higher Education Commission’s National Research Program for Universities (NRPU) under project number 20-15756 and title \textquotesingle Generating Plausible Counterfactual Explanations using Transformers in Natural Language-based Financial Forecasting (NLFF)\textquotesingle.

\section*{Acknowledgements}
We are deeply grateful for HEC-NRPU financial support, which allowed us to conduct the experiments and analysis presented in this paper. This support has facilitated collaboration and knowledge-sharing among our research team and Falki Capital (Pvt) Ltd. (Equities brokerage house). The authors also acknowledge the invaluable support and resources provided by the CPInS Lab at SEECS-NUST, which has been instrumental in facilitating the research and publication of this work

\bibliography{sn-bibliography}

\section*{Figure Legends}
Fig.1 KSE-100 ten years closing index\\Fig.2 EMA 55, 100 and 200 of KSE-100 index stocks\\
Fig.3 Potential trading signals from EMA-55 strategy\\
Fig.4 Potential trading signals from RSI strategy\\
Fig.5 Visualization of feature importance\\
Fig.5(a) 1-day trading\\
Fig.5(b) Weekly trading\\
Fig.6 Block plot visualization for proposed trading strategies\\
Fig.6(a) Daily trading\\
Fig.6(b) Short-term trading\\
Fig.7 Force plot visualization for the proposed short-term trading strategy instances\\
Fig.7(a) 'hold a share' trading move\\
Fig.7(b) 'buy a share'trading move\\
Fig.8 Decision plot indicating significant contributors in decision pattern using machine
learning model explanation\\
Fig.9 Force plot indicating significant contributors using machine learning model explanation\\
Fig.10 Investment returns for a conservative portfolio\\
Fig.11 Investment returns for an aggressive portfolio

\end{document}